\documentclass[12pt]{article}
\usepackage{epsfig}
\usepackage{graphicx}
\def\slash#1{\setbox0=\hbox{$#1$}#1\hskip-\wd0\dimen0=5pt\advance
       \dimen0 by-\ht0\advance\dimen0 by\dp0\lower0.5\dimen0\hbox
         to\wd0{\hss\sl/\/\hss}}

\newcommand{\be}{\begin{equation}}
\newcommand{\dd}{\displaystyle}
\newcommand{\ee}{\end{equation}}
\newcommand{\bea}{\begin{eqnarray}}
\newcommand{\eea}{\end{eqnarray}}

\begin{document}
\hfill{\bf BARI-TH 444/02}\par \hfill{\bf FIRENZE DFF
393/08/02}\par \hfill{\bf UGVA-DPT-2002-09/1102}
\begin{center}
{\Large{\bf{Massive quark effects in two flavor color
superconductors}}}
\end{center}
\begin{center}
{\Large\bf\boldmath {}} \rm \vskip1pc {\large R.
Casalbuoni$^{a,b}$, F. De Fazio$^{d}$, R. Gatto$^e$,
  G. Nardulli$^{c,d}$ and M. Ruggieri$^{c,d}$}\\ \vspace{5mm} {\it{
$^a$Dipartimento di
 Fisica, Universit\`a di Firenze, I-50125 Firenze, Italia
 \\
 $^b$I.N.F.N., Sezione di Firenze, I-50125 Firenze, Italia\\ $\dd
^c$Dipartimento di Fisica,
 Universit\`a di Bari, I-70124 Bari, Italia  \\$^d$
 I.N.F.N.,
 Sezione di Bari, I-70124 Bari, Italia\\
 $^e$D\'epart. de Physique Th\'eorique, Universit\'e de Gen\`eve,
 CH-1211 Gen\`eve 4, Suisse }}
 \end{center}
  \begin{abstract}
The high density effective theory formalism (HDET) is employed to
describe high density QCD with  two massive flavors (2SC).  The
gap equation is derived and explicitly solved for the gap
parameter. The parameters associated to the pseudo Nambu-Goldstone
boson of $U(1)_A$ are evaluated in the limit $\mu\to\infty$ and
$m/\mu$ fixed. In particular we find for the velocity of the NG
boson the relation
$v^2=\sqrt{\mu_1^2-m_1^2}\sqrt{\mu_2^2-m_2^2}/3\mu_1\mu_2$.
 \end{abstract}

\section{Introduction}
In the present letter we  study the effect of quark masses in the
two flavor color superconductor \cite{Alford:1997zt} (for a
recent review see \cite{Rajagopal:2000wf}) in a BCS pairing.
Correspondingly we will take the masses of two quarks not too
different from each other.  Massive quarks have been studied by
several authors both for two and  three flavors
\cite{Son:1999cm}, \cite{Rho:1999xf}, \cite{Hong:1999ei},
\cite{beane}, \cite{raku}, \cite{Schafer2}. In most of the papers
quark masses are treated perturbatively, under the hypothesis
that they are much smaller than the energy gap. However this
assumption is questionable for the strange quark, which may affect
phenomenological analyses (for examples see
\cite{Bedaque:2001je}, \cite{Kaplan:2001qk}). In the present
paper we make a first attempt to go beyond the above mentioned
approximation by considering the limit $\mu\to\infty$ with
$x\equiv m/\mu$ fixed. We will perform our analysis within the
computational scheme of the High Density Effective Theory (HDET),
\cite{HDET,casa2} which is most suitable for calculations
involving the high density limit of QCD. We will limit most of
our analysis to the case of two flavors, though an extension to
three flavors is in principle admissible. In Section 2 we
establish our notations.  Section
 3 is dedicated to the 2SC model with massive quarks where
we  study the gap equation. These results generalize the results
of \cite{raku}. In Section 4 we analyze the properties of  the
pseudo Nambu Goldstone Boson (NGB) related to the spontaneous
breaking of the $U(1)_A$ symmetry. We recover known results for
the mass, although with a different method. On the other hand we
find a new  mass effect in the velocity of the pseudo NGB, which
is given by \be
v^2=\frac{\sqrt{\mu_1^2-m^2_1}\sqrt{\mu_2^2-m^2_2}}{3\mu_1\mu_2}\
,\ee where $\mu_k,m_k$ ($k=1,2$) are the masses and quark chemical
potentials of the two quarks in the Cooper pair. This result is
not in contradiction with the findings of other authors, see e.g.
\cite{beane}, since our corrections are of the order $m/\mu$
rather than $m/\Delta$. We also argue that this result should be
true at all orders in our parameter expansion.

\section{The effective lagrangian for a massive gapless quark}
Starting point is the QCD lagrangian \be {\cal
L}_{QCD}=\bar{\psi}\,(i\,\slash D+\mu\,\gamma^{0}-m)\,\psi=
\bar{\psi}\,(i\,\slash D +\mu\,\gamma^{0}-x\mu)\,\psi\label{lag}
\ee where we shall held \be x=\frac m\mu\ ,\hskip1cm 0<x<1
\label{x}\ee fixed in the $\mu\to\infty$ limit.

 We introduce velocity dependent  fields $\psi_{\vec v}(y)$ and $\Psi_{\vec
 v}(y)$\be
\psi_{\vec v}(y)={\cal P}_{+}
\int_{|\ell\,|<\delta}\frac{d^{4}\ell}{(2\,\pi)^{4}}\,e^{i\,\ell\cdot
y}\,\tilde\psi(\ell),~~ \Psi_{\vec v}(y)={\cal P}_{-}
\int_{|\ell\,|<\delta}\frac{d^{4}\ell}{(2\,\pi)^{4}}\,e^{i\,\ell\cdot
y}\,\tilde\psi(\ell)\ ,\label{decompose}
 \ee where $\Delta\ll\delta\ll\mu$ and we have written the quark
 momentum as $ p^\mu=\mu v^\mu+\ell^\mu \ ,\label{dec}$
  with $v^\mu=(0,\vec v)$. In (\ref{decompose}) we have used the projectors
  \begin{equation}
{\cal
P}_{\pm}=\frac{1\pm(\vec{\alpha}\cdot\vec{v}+x\,\gamma^{0})}{2}
\label{energyprojectors1} \ .
\end{equation} These projectors can be obtained from the
conventional ones, $(1\pm\vec\alpha\cdot\vec v)/2$, via a
Cini-Touschek transformation \cite{cini}. We recall that this
transformations was introduced to study the ultra-relativistic
limit of the Dirac equation. Owing to (\ref{decompose}) we can
write \be\psi(y)= \sum_{\vec n} e^{-i\mu v\cdot y}
 \left[\psi_{\vec v}(y)+\Psi_{\vec v}(y)\right]\ ,\label{eq.5}\ee
where the sum represents an average over the velocity directions,
with $\vec n=\vec v/|\vec v|$ and arises because there are
infinitely many ways to decompose the quark momentum according to
(\ref{dec}). $\psi_{\vec v}(y)$ and $\Psi_{\vec v}(y)$ represent
respectively the positive energy and negative energy solution of
the Dirac equation associated to the lagrangian (\ref{lag}).
 We now substitute (\ref{eq.5})
 in ${\cal L}_{QCD}$ and make use of the Riemann-Lebesgue lemma to eliminate one of the two sums
 appearing in the lagrangian (velocity superselection rule of HDET); therefore we
 get
 \begin{equation}
{\cal L}_0=\sum_{\vec{n}}\left\{\psi^{\dagger}_{\vec
v}\,i\,\gamma^{0}\slash D\,\psi_{\vec v} +\Psi^{\dagger}_{\vec
v}(i\,\gamma^{0}\slash D+2\mu)\,\Psi_{\vec v}
+\Psi^{\dagger}_{\vec v}\,i\,\gamma^{0}\slash D\,\psi_{\vec v}+
\psi^{\dagger}_{\vec v}\,i\,\gamma^{0}\slash D\Psi_{\vec v}
\right\} \label{intermediate2}
\end{equation}
For quarks near the Fermi surface  we can write $|\vec p|\approx
\mu |\vec v|$ and $m^{2}=\mu^{2}\,(1-v^{2})$; therefore the
parameter $x$ defined in (\ref{x}) is also given by \be
x=\sqrt{1-v^{2}} \ .\label{x1}\ee  We will use the identities
\bea \bar{\psi}_{\vec v}\,\gamma^{0}\,{\psi}_{\vec
v}={\psi}_{\vec v}^{\dagger}\, {\psi}_{\vec v},&&~~~~~~
\bar{\psi}_{\vec v}\,\vec{\gamma}\,{\psi}_{\vec v} ={\psi}_{\vec
v}^{\dagger}\,\vec{v}\,{\psi}_{\vec v}\ ,\cr&&\cr \bar{\Psi}_{\vec
v}\,\gamma^{0}\,\Psi_{\vec v}=\Psi^{\dagger}_{\vec v}\,
{\Psi}_{\vec v},&&~~~~~~ \bar{\Psi}_{\vec
v}\,\vec{\gamma}\,\Psi_{\vec v}=-\Psi^{\dagger}_{\vec
v}\,\vec{v}\,\Psi_{\vec v}\  ,\eea and  \be {\cal
P}_-\,\vec\alpha\cdot\vec v\, {\cal P}_+=x(x-\gamma_0)P_+\
.\label{11}\ee  We can integrate out in the functional
integration $\Psi_{\vec v}$ through the equation \be \Psi_{\vec
v}=\frac{-i\,\gamma^{0}\slash D}{i\,\tilde{V}\cdot
D+2\mu}\,\psi_{\vec v}\ , \label{!!!}\ee where we have used the
notations $V^{\mu}=(1,\vec{v}),\tilde{V}^{\mu}=(1,-\vec{v})$

 As in the case of massless flavor,
the $\Psi_{\vec v}$ fields are of order $1/\mu$ with respect to
$\psi_{\vec v}$ fields; therefore, at the leading order in
$1/\mu$ one can write the Dirac part of the massive quark
lagrangian as
\begin{equation}
{\cal L}_0=\sum_{\vec{n}}\,\psi^{\dagger}_{\vec v}\,i\,V\cdot
D\,\psi_{\vec v}\ . \label{leading1}
\end{equation}

The  integration of the antiquark fields gives a correction to the
kinetic term of order $x^2/\mu$. This correction can be neglected
in our limit.

Since the sum over velocities is symmetric over the entire solid
angle, one can write ${\cal L}_0$ as follows:
\begin{equation}
{\cal L}_0=\,\frac{1}{2}\,\sum_{\vec{n}}\{
\psi^{\dagger}_{+}\,(i\,V\cdot D)\,\psi_{+}+
\psi^{\dagger}_{-}\,(i\,\tilde{V}\cdot D)\,\psi_{-}\}
\label{lagrKN2}
\end{equation}
 where the subscripts $\pm$ denote the
velocity direction, i.e. \be \psi_{\pm}=\psi_{\pm\vec v}
\label{twocomponentspinor1bis} \ . \ee This equation is formally
the same as in the massless case, however one should remember
that the velocity satisfies the condition (\ref{x1}). Another
important difference is that we do not separate left and right
degrees of freedom, and $\psi_{\pm}$ have two independent
components. Finally
 the average over the directions is defined as follows:\be \sum_{\vec
n}=\int\frac{d\vec n}{8\pi}\ .\label{sumvel} \ee  The extra factor
$1/2$ in the average is introduced because, after the introduction
of the field with opposite velocity $\psi_-$, one doubles the
degrees of freedom, which implies that the integration must be
only over half solid angle. For the following it will be useful
to introduce the Nambu-Gorkov notation \be
\chi=\frac{1}{\sqrt{2}}\left(\matrix{\psi_+\cr
C\psi_{-}^*}\right)\,.\label{chi}\ee

\section{2SC with massive quarks}

We consider the model with two massive quarks of different flavor
in a color superconductor state \cite{raku}. At the Fermi surface
one has \be \vec p_1=\mu_1 \vec v_1,~~ \vec p_2=\mu_2 \vec v_2
\ee Let us put \be \mu_{1}= \mu-\delta\mu,~~~
\mu_{2}=\mu+\delta\mu\,. \ee The Cooper pair in the 2SC model
couples quarks of different flavor, and the total momentum of the
pair must vanish \be \vec p_1+\vec p_2=0\ .\label{pp}\ee One gets
immediately, from $p_{1}=p_{2}$ that
\begin{equation}
\frac{\delta\mu}{\mu}=\frac{\sqrt{1-x_1^{2}}-
\sqrt{1-x_2^{2}}}{\sqrt{1-x_1^{2}}+\sqrt{1-x_2^{2}}}=\frac{m_2^2-m_1^2}{4\mu^2}\,
, \label{dmu1}
\end{equation} where $x_{i}=m_{i}/\mu_{i}$.
Eq. (\ref{dmu1}) is a consequence of (\ref{pp}) and fixes
$\delta\mu$ once the masses and $\mu$ are given. Here we are
taking advantage of the fact that only strong interactions are
considered and therefore $\delta\mu$ has no constraints. By
including also weak interactions, this line of reasoning might
fail. As discussed in \cite{raku}, however, if the deviations from
(\ref{pp}) and (\ref{dmu1}) are small, the value of the gap
remains inaltered. Only for larger deviations one has a change,
since there is a phase transition to the crystalline
superconducting state \cite{Alford:2000ze}, \cite{Bowers:2002xr}.
Eventually, for very large deviations, the normal non
supercondcting state becomes energetically favorite.

In order to write down the HDET lagrangian we first observe that
the Dirac lagrangian not including the gap terms is simply given
by two terms (\ref{lagrKN2}), one for each flavor. In the 2SC
phase we have
 \be \langle 0|\psi^T_{\alpha
i}\,C\,\gamma_5\psi_{\beta j}|0\rangle=
\Delta\,\epsilon^{\alpha\beta 3}\,\epsilon_{ij 3}\neq 0\
,\label{cond} \ee where $i$ and $j$ are the flavor indices and
$\alpha$ and $\beta$ are color indices. Therefore we add the term
\be {\cal L}_{gap} ={\cal L}_\Delta+ {\cal L}_{\bar \Delta} \,,\ee
with \be {\cal L}_\Delta=-\sum_{\vec n}\frac\Delta 2
\epsilon^{\alpha\beta 3}\,\epsilon_{ij 3}\psi^{T}_{\alpha i,-}
\,C\,\gamma_5\psi_{\beta j,+}+h.c.\ee and\be{\cal L}_{\bar
\Delta}= -\frac\Delta 2 \epsilon^{\alpha\beta 3}\,\epsilon_{ij
3}\sum_{\vec n} \Big(\Psi^{T}_{\alpha
i,-}\,C\gamma_5\,\Psi_{\beta j,+} + \psi^{T}_{\alpha
i,-}\,C\,\gamma_5\Psi_{\beta j,+}+\Psi^{T}_{\alpha
i,-}\,C\,\gamma_5\psi_{\beta j,+}\Big)+ h.c.
 \label{antigap2}
\ee ${\cal L}_{\bar \Delta}$ contains antiquark operators. These
operators are formally suppressed but, as in the case of the
Meissner mass \cite{casa2}, they may give rise to divergent
contributions. This happens for the mass of the $U(1)_A$ boson,
where we get an unsuppressed term at the leading order in $\mu$.
We notice that by the equation (\ref{antigap2}) we implicitly
assume that the antiquark gap $\bar\Delta$ is equal to the quark
gap $\Delta$. Using eqs. (\ref{!!!}) and (\ref{11}) we can write
in momentum space
 for free quarks
\begin{equation}
\Psi_{\vec v}= \frac{-x (x-\gamma_0)\ell_\parallel}{|\vec v|(2\mu
+\tilde V\cdot \ell)}\psi_{\vec v}\,,\end{equation} which is
proportional to the quark mass.

This introduces in the fermionic lagrangian in the 2SC model the
non local interaction term
\begin{eqnarray}
{\cal L}_{\bar{\Delta}}&=&-\frac{\Delta}{2}\, \epsilon^{\alpha
\beta 3}\,\epsilon_{i j 3}\, \sum_{\vec{n}} \psi^{T}_{\alpha
i,-}\, C\,\Big\{ \frac{\ell_\parallel x_i(x_i+\gamma_0)}{|\vec
v_i|(2\,\mu_i+V_{i}\cdot \ell)}\gamma_5 \frac{\ell_\parallel
x_j(x_j-\gamma_0)}{|\vec v_j|(2\,\mu_j+\tilde V_{j}\cdot
\ell)}\cr&&\cr&-& \gamma_5 \frac{\ell_\parallel
x_j(x_j-\gamma_0)}{|\vec v_j|(2\,\mu_j+\tilde V_{j}\cdot \ell)}-
\gamma_5\frac{\ell_\parallel x_i(x_i-\gamma_0)}{|\vec
v_i|(2\,\mu_i+V_{i}\cdot \ell)}\Big\} \psi_{\beta j,+}
\,+\,h.c.\label{48}
\end{eqnarray}
These terms will be taken  in the large $\mu$ limit. Since
 ${\cal L}_{\bar\Delta}$ can  contribute, in the
leading order in $\mu$,  only to divergent diagrams (i.e. giving
rise to $\mu$ factors), it  may be treated  as an insertion.

Let us introduce \be
\chi_\alpha=\left(\matrix{\chi_{\alpha,1}\cr\chi_{\alpha,2}}\right)\,,\ee
where $\chi_{\alpha,i}$ are the Nambu-Gorkov fields (see eq.
(\ref{chi})) for the two flavors. Then we get: \bea {\cal
L}=\sum_{\vec{n},\alpha,\beta}\, \chi^{\dagger}_{\alpha}
\left(\matrix{ iV_{1}\cdot D^{\alpha\beta} &0 &0 &
\hat\Delta^{\alpha\beta}\cr 0 & i\tilde V_{1}\cdot
D^{*,\alpha\beta} &\hat\Delta^{\alpha\beta} & 0\cr 0&
-\hat\Delta^{\alpha\beta} & iV_{2}\cdot D^{\alpha\beta} &0 \cr
-\hat\Delta^{\alpha\beta} & 0& 0& i\tilde V_{2}\cdot
D^{*,\alpha\beta}}\right)\chi_{\beta}\,,\eea \be
\hat\Delta^{\alpha\beta}=\gamma_5\Delta\,\epsilon^{\alpha\beta 3}
\ .\label{twoflavorgapmatrix1} \label{deff1} \ee

To get the quark propagator we define \be
D_{1}(\ell,\Delta)=V_{1}\cdot \ell\, \tilde{V}_{2}\cdot \ell-
\Delta^{2},~~~~ D_{2}(\ell,\Delta)=V_{2}\cdot \ell\,
\tilde{V}_{1}\cdot \ell- \Delta^{2}\, . \ee We get \be
S^{\alpha\beta}(\ell)= \left(\matrix{\dd{ \frac{\tilde{V_{2}}\cdot
\ell}{D_{1}}\delta^{\alpha\beta}}& 0& 0&
\dd{-\frac{\hat\Delta^{\alpha\beta}}{D_{1}}}\cr 0&
\dd{\frac{{V_{2}}\cdot \ell}{D_{2}}\delta^{\alpha\beta}}&
\dd{-\frac{\hat\Delta^{\alpha\beta}}{D_{2}}}&0\cr
0&\dd{\frac{\hat\Delta^{\alpha\beta}}{D_{2}}}& \dd{\frac{\tilde
V_{1}\cdot \ell}{D_{2}}\delta^{\alpha\beta}}&0\cr
\dd{\frac{\hat\Delta^{\alpha\beta}}{D_{1}}}&0&0&
\dd{\frac{V_{1}\cdot \ell}{D_{1}}\delta^{\alpha\beta}} }\right)
\label{fermionicpropagator}\,. \ee To get the gap equation in the
2SC model we use the Schwinger-Dyson equation with a four-fermion
interaction:
\begin{equation}
\Sigma^{\alpha\beta}= -ig^{2}\int\frac{d^{4}k}{(2\pi)^{4}}
\,\,\frac{g_{\mu\nu}\,\delta^{a
b}}{\Lambda^{2}}\,\Gamma_{a}^{\mu,\alpha\gamma}
\,S_{\gamma\delta}(k)\,\Gamma_{b}^{\nu,\delta\beta}\,,
\label{scwdy1}
\end{equation}
where $\Sigma=-[S^{-1}-S^{-1}_{F}]$,  $S^{-1}_{F}$ stands for the
inverse of the free quark propagator and\begin{displaymath}
\Gamma_a^{\mu,\alpha\beta}= \left(\matrix{ i\,V^{\mu}_{1}\,\cdot
T_{a}^{\alpha\beta} &&& \cr &-i\,\tilde V^{\mu}_{1}\,\cdot
T_{a}^{T,\alpha\beta}&&\cr && \,i\,{V}^{\mu}_{2}\,\cdot
T_{a}^{\alpha\beta}&\cr &&& -\,i\,\tilde{V}^{\mu}_{2}\,\cdot
T_{a}^{T,\alpha\beta} }\right)\,;
\end{displaymath}
$T_{a}^{\alpha\beta}$ being the usual $SU(3)$ generators. We get:
\begin{equation}
\hat\Delta^{\alpha\beta}=-i\,\frac{g^{2}}{\Lambda^{2}}\,\tilde{V_2}\cdot
V_{1}\,(T_a\hat\Delta T^T_a)^{\alpha\beta}
\,4\pi(\mu^2-\delta\mu^2)\,\alpha(x_1,x_2)\int\,
\frac{d^2\ell}{(2\pi)^{4}}\frac{1}{D_{1}(\ell,\Delta)}
\label{twoflavorgapeq1.0}
\end{equation}
The factor $(\mu^2-\delta\mu^2)\,\alpha(x_1,x_2)=\mu_1^2|\vec
v_1|^2=\mu_2^2|\vec v_2|^2=p_F^2$ implies a reduction of the
phase space with respect to the massless case. For the sequel the
following notations will be useful: \be
\alpha(x_1,x_2)=\sqrt{1-x_1^{2}}\,\sqrt{1-x_2^{2}},~~~
\beta(x_1,x_2)=\sqrt{1-x_2^{2}}-\sqrt{1-x_1^{2}}\, .\ee The gap
equation, for $\Delta\neq 0$ is
\begin{equation}
1=i\,\frac{g^{2}(\mu^2-\delta\mu^2)\alpha(x_1,x_2)}{3\Lambda^{2}\pi^3}\frac{\tilde{V_2}\cdot
V_{1}}2 I(\Delta,x_1,x_2)\,,\ee where \be I(\Delta,x_1,x_2)=
\int\,\frac{d^2\ell}{\ell_{0}^{2}-\alpha(x_1,x_2)
\ell_{\parallel}^{2}+\beta(x_1,x_2)\ell_{0}
\ell_{\parallel}-\Delta^2 }\ . \label{twoflavorgapeq1} \ee where
the limits of $\ell_\parallel$ integration are $\pm\delta$.
 One can easily check that for $x_1=x_2=0$ one recovers the gap
equation for the 2SC model in the HDET approximation and for
massless quarks\footnote{See eqs. (2.188) and (2.184) of
\cite{nard1}.}. Performing the integral we obtain:\be
\Delta=\sqrt{\frac{4\alpha+\beta^2}{4}}\, \frac{\delta}{\dd
\sinh\left(\frac{3\Lambda^2\pi^2}{(\mu^2-\delta\mu^2)\alpha
g^2}\,\frac{\sqrt{ 4\alpha+\beta^2}}{2(1+\alpha)
}\right)}\label{complete}\,.\ee If the flavor $1$ means the quark
up and $2$ the strange quark we can write  $x_1\approx 0;\,
x\equiv x_{2}$ and approximate (\ref{dmu1}) as follows:
 \begin{equation}
\frac{\delta\mu}{\mu}=\frac{1-
\sqrt{1-x^{2}}}{1+\sqrt{1-x^{2}}}\,, \label{dmu1bis}
\end{equation}or
\be \frac{\delta\mu}{\mu}=\frac{m_2^2}{4\mu^2}\,,\label{38bis}\ee
\ and we get \be
\Delta\approx\left(1-\frac{\delta\mu}{\mu}\right)\frac{\delta}{\dd
\sinh\left(\frac{3\Lambda^2\pi^2}{2\mu^2g^2(1-\frac{2\delta\mu}\mu)
}\right)}\label{approx}\,.\ee

\begin{figure}[htb]
\epsfxsize=8.5truecm \centerline{\epsffile{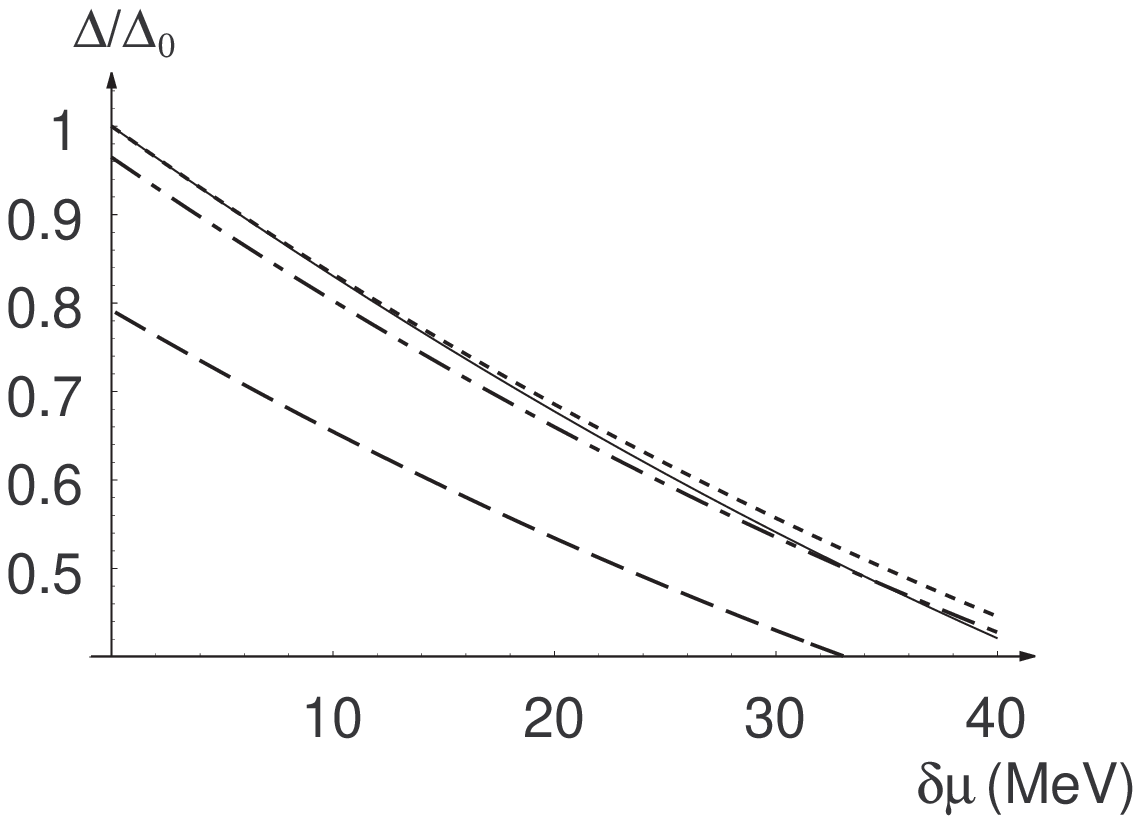}}
\label{fig2} Fig. 1 The ratio of the gap for finite masses
$\Delta$ over the gap for massless quarks $\Delta_0$ as a
function of $\delta\mu$. The solid line refers to the approximate
expression for $\Delta$, see eq. (\ref{approx}), whereas the
other three refer to the complete expression in eq.
(\ref{complete}) with the dotted line corresponding to $x_1=0$,
the dash-dotted line to $x_1=0.1$, and the dashed one to
$x_1=0.25$. The values used are $\mu=400$ MeV,
 $\Lambda/g=181$ MeV, corresponding to $\Delta_0\approx 40~MeV$.
 Notice that plotting $\delta\mu$ is  the same as plotting
 the squared masses difference, $m_2^2-m_1^2$ due to the
 relation (\ref{dmu1}).
\end{figure}
In Fig. 1 we show the ratio of the gap for finite masses $\Delta$
over the gap for massless quarks $\Delta_0$ as a function of
$\delta\mu$. In the plot $x_2$ has been traded off for
$\delta\mu$ and different values of $x_1$ have been used. Also
the difference between the approximate expression and the exact
one is given for $x_1=0$.

The greatest effect of the strange quark mass on the condensation
energy of the BCS pair is due to the reduction of the size of the
Fermi surface. This implies that for small values of $\delta\mu$
the condensation energy of the pair decreases linearly with
$\delta\mu$.  The $x_1=0$ curve (dotted line) plots the same
quantity as in ref. \cite{raku} and agrees quantitatively with the
results found there. It can be observed that also the results of
\cite{raku}, similarly to what we have done here, are neither
obtained by expanding in $m/\Delta$ nor in $m/\mu$. The
differences between our calculation and that of \cite{raku} are as
follows. Our eq. (37) presents the advantage to offer an
analytical expression; moreover it can be used when both masses,
and not only one,  are nonvanishing. On the other hand our
results are only valid for $m_i\le\mu$; for higher values the
evaluations of \cite{raku}, that do not suffer of this limitation,
should be used.

We note that the insertion of ${\cal L}_{\bar\Delta}$ would not
change the main result we found, mainly the reduction of the
phase space, because its effects would only change $\Delta$ by
factors of order $x/\mu$.

\section{$U(1)_{A}$ pseudo Nambu-Goldstone mode} Besides color
symmetry, that is broken from $SU(3)_c$ down to $SU(2)_c$, the
condensate (\ref{cond}) also breaks $U(1)_B$ and $U(1)_{em}$;
however two linear combinations of these abelian subgroups remain
unbroken and there are no NGBs associated to them. The vacuum
expectation value (\ref{cond})  breaks spontaneously $U(1)_A$,
which is however also broken by the strong anomaly. At high $\mu$
this latter breaking is soft and one expects that the associated
pseudo Goldstone boson, the $\eta^\prime$, is almost massless. In
the sequel we will study the effect on $\eta^\prime$ of the quark
masses in the approximation of neglecting  the strong anomaly.

We introduce an external field $\sigma$ associated to the pseudo
Goldstone boson and we put
\begin{displaymath}
U=\exp\left\{i\,\frac{\sigma}{f_{\sigma}}\right\}\,.
\end{displaymath} In (\ref{chi}) we perform the substitution
\be \psi_{\alpha_j+}\rightarrow U^\dagger \psi_{\alpha_j+},~~~~
\psi^*_{\alpha_j-}\rightarrow U \psi^*_{\alpha_j-}
\label{67}\,,\ee
 and ${\cal
L}_{0}$+${\cal L}_{\Delta}$ becomes ${\cal L}_{0}$ +${\cal
L}_{\Delta+\sigma}$, with
\begin{equation}
{\cal L}_{\Delta+\sigma}=\sum_{\vec{n}}
\chi^{\dagger}_{\alpha}\,\hat\Delta^{\alpha\beta} \left(\matrix{
 & & & U^{2}\cr
 & & U^{\dagger\, 2} & \cr
 & -U^{2} & & \cr
 -U^{\dagger\,2}& & &
}\right)\,\chi_{\beta}\ . \label{goldston1}
\end{equation}
At the second order in the $\sigma$ field: \be U^{2}\approx
1+\frac{2 i \sigma}{f_{\sigma}}-\frac{2\sigma^{2}}{f^{2}_{\sigma}}
\ee and the quark-boson interaction lagrangian can be written as
follows:
\begin{equation}
{\cal L}_{\sigma}=\frac{2i\sigma}{f_\sigma}\sum_{\vec{n}}
\chi^{\dagger}_{\alpha}\hat\Delta^{\alpha\beta}\Gamma_{0}\chi_{\beta}
-\frac{2\sigma^2}{f^2_\sigma}\sum_{\vec{n}}
\chi^{\dagger}_{\alpha}\hat\Delta^{\alpha\beta}\Gamma_{1}\chi_{\beta}\label{goldston2}\,,
\end{equation}
where \be \Gamma_{0}=\left(\matrix{  &&&1\cr &&-1&\cr &-1&&\cr
1&&& }\right)~,~~~~~ \Gamma_{1}=\left(\matrix{ &&&1\cr &&1&\cr
&-1&&\cr -1&&& }\right)\ .\ee The $\sigma$ kinetic term  arises
from  loop expansion after functional integration of the
fermionic degrees of freedom. Therefore we have  the following
effective action
\begin{eqnarray}
{\cal S}_{eff}&=& i\, \frac{1}{2}\,{\rm{Tr}} \int dx\,dy\, \left\{
i\,S(y,x)\,\frac{2\,i\,\sigma(x)}{f_{\sigma}}{\hat\Delta}\,i\,\Gamma_{0}
\,i\,S(x,y)\,\frac{2\,i\,\sigma(y)}{f_{\sigma}}{\hat\Delta}\,i\,\Gamma_{0}
\right\} \cr &+&i{\rm{Tr}}\int
dx\,\left\{i\,S(x,x)\,\frac{(-2)}{f^{2}_{\sigma}}
{\hat\Delta}\,\sigma^{2}(x)\,i\,\Gamma_{1} \right\}\,.
\label{loopexpansion5}
\end{eqnarray}In evaluating the trace over the spin indices one has to include
the ${\cal P}_+$ projector which gives a factor 2 if no other
$\gamma$-matrices are involved. The two terms correspond to the
self-energy and a tadpole diagram respectively, see fig. 2.

\begin{figure}[ht]
\epsfxsize=8truecm \centerline{\epsffile{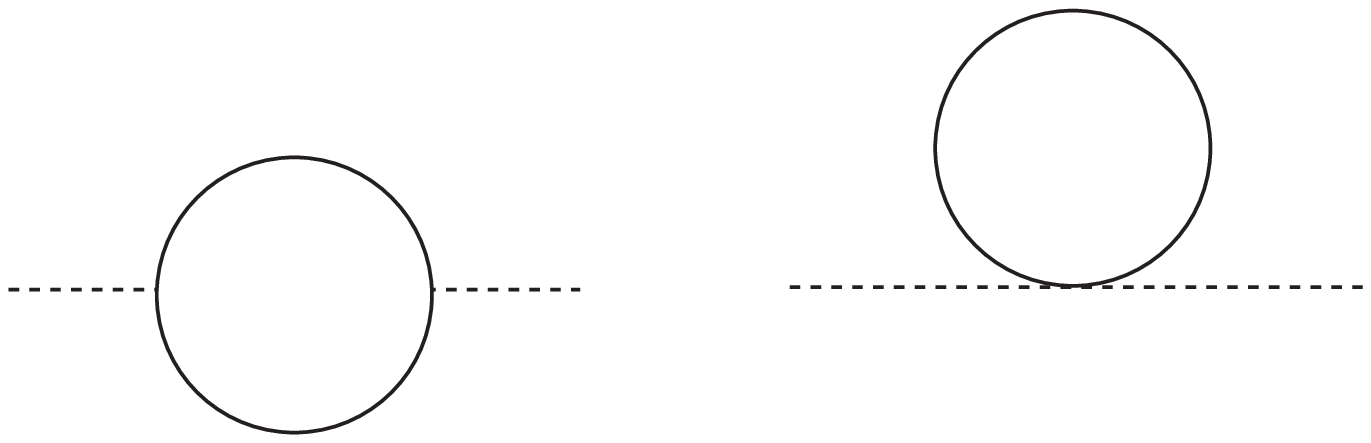}} \noindent
Fig. 2 One-loop diagrams. External lines represent the pseudo
Goldstone field. Full lines are fermion propagators.
\end{figure}

The result of the calculation of the effective lagrangian in
momentum space is as follows: \bea &&i\,{\cal
L}_{s.e.}(p)=\frac{16\,\mu_1\mu_2\,\alpha\,\Delta^{2}\,
}{8\pi^{3}\,f^{2}_{\sigma}}\,\sum_{\vec{n}}\,\int d^{2}\ell
\cr&&\cr&& \Big[ \frac{V_{1}\cdot
\ell\sigma\,\tilde{V}_{2}\cdot(\ell+p)\sigma+V_{1}\cdot(\ell+p)
\sigma\,\tilde{V}_{2}\cdot
\ell\sigma-2\,\Delta^{2}\sigma^2}{D_{1}(\ell)\,D_{1}(\ell+p)}
+1\longleftrightarrow 2 \Big]\label{selfenergy1}\eea \be i\,{\cal
L}_{tad}=-\frac{4\,\mu_1\mu_2\,\alpha\,\Delta^{2}\,\sigma^{2}}{\pi^{3}
\,f_{\sigma}^{2}} \,\,\sum_{\vec{n}}\,\int
d^{2}\ell\left(\frac{1}{D_{1}(\ell,\Delta)}+\frac{1}{D_{2}(\ell,\Delta)}\right)
\label{tadpole1} \,.\ee We find
 \be{\cal L}_{s.e.}(p=0)+{\cal L}_{tad}=0\ .\ee

We have to add the contribution from the antiquark gaps. The
pseudo NGB masses in color superconducting phases  take
contribution from different sources. In the case of the singlet
pseudo NGB, as shown in \cite{Schafer2}, its mass arises from an
effective four quark operator obtained by integrating out the
electric gluon field; the result is therefore
proportional\footnote{The calculations in \cite{Schafer2} are in
the Color Flavor Locking phase and should be adapted to the 2SC
case.} to $m_1m_2\Delta^2/\mu^2$. In this approach the antiquark
gap terms do not appear explicitly  as the antiquark field
operators are integrated out.

Also in the present approach we integrate out the antiquark
operators, as discussed above and ${\cal L}_{\bar \Delta}$ is
expressed in terms of the quark fields only. It contributes both
to ${\cal L}_{s.e.}(0)$ and to ${\cal L}_{tad}$, through two
additional vertices $\sigma\chi\chi$ and $\sigma\sigma\chi\chi$
that can be constructed analogously to (\ref{goldston2}) using
(\ref{48}) and (\ref{67}). We insert these vertices in the
expressions for the self-energy and for the tadpole once, since
we will take only the leading term in the quark masses to the
mass of the Goldstone. The two contributions do not cancel out
and one finds \be i{\cal
L}_{eff}(0)=\frac{4\Delta^2\alpha\sigma^2}{\pi^3f^2_\sigma}x_1x_2\hat
I\ . \label{m}\ee  For $\hat I$ we get:
\begin{displaymath} \hat I=
\int_{-\mu}^{+\mu}d^2\ell\,\frac{(\mu+\ell_\parallel)^2
\,\ell_\parallel^2}{(2\mu+\tilde V\cdot\ell)(2\mu+
V\cdot\ell)D(\ell,\Delta)} \approx-2\pi
i\mu^2\,\log\frac\mu\Delta\,+{\cal O}(\mu^2)\,. \end{displaymath}
This is the leading contribution and arises from the first term in
(\ref{48}), since the other two terms are of order $\mu^2$ and not
$\mu^2\log\mu$.

Let us then consider ${\cal L}_{eff}(p)-{\cal L}_{eff}(0)$.
\begin{eqnarray}
&&i\,{\cal L}_{eff}(p)-i\,{\cal L}_{eff}(0)
=\frac{4\,\mu_1\mu_2\,\alpha\,\Delta^{4}}{\pi^{3}\,f^{2}_{\sigma}}
\,
 \sum_{\vec{n}}(V_1\cdot p)\sigma\,(\tilde V_2\cdot p)\sigma\,
 \int\,\frac{d^{2}\ell}{[D_{1}(\ell)]^3}\,+\cr&&\cr&&+
 \frac{2\,\mu_1\mu_2\,\alpha\,\Delta^{4}}{\pi^{3}\,f^{2}_{\sigma}}\,\,
\sum_{\vec{n}}\,\int\,\frac{d^{2}\ell}{[D_{1}(\ell)]^3}\times\cr&&\times\left(
 [(\tilde V_2\cdot p)\sigma]^2\,[
  V_1\cdot \ell]^2\,+\,[(V_1\cdot p)\sigma]^2\,\left(
 \tilde V_2\cdot \ell\right)^2\right)+\left(1\leftrightarrow
 2\right)
\label{selfenergy10} \,.\eea
 One can show that the second term on the r.h.s. of
(\ref{selfenergy10}) vanishes, while one finds \be
\int\,\frac{d^{2}\ell}{[D_{1}(\ell)]^3}=
\frac{-\,\pi\,i}{\Delta^4\sqrt{4\alpha+\beta^2}}\,.
 \ee
 We get therefore
\begin{displaymath}
{\cal
L}_{eff}(p)=\,-\,\frac{4\,\mu_1\mu_2\,\alpha}{\pi^{2}\,f^{2}_{\sigma}}
\,\frac{1}{\left(4\alpha+\beta^2\right)^{1/2}}\,
 \sum_{\vec{n}}(\tilde{V}_{2}^{\mu}\,V_{1}^{\nu}+\tilde{V}_{1}^{\mu}\,V_{2}^{\nu})\,
p_{\mu}\,\sigma\,p_{\nu}\,\sigma\,.
\end{displaymath}
From which one finally gets, in coordinate space
\begin{displaymath}
{\cal
L}_{eff}(\sigma)=\frac{4\,\mu_1\mu_2\,\alpha}{\pi^{2}\,f^{2}_{\sigma}}
\,\frac{1}{\left(4\alpha+\beta^2\right)^{1/2}}\, \left(
(\partial_{0}\,\sigma)^{2}-\frac{\alpha}{3}(\vec{\partial}\,\sigma)^{2}-
m_\sigma^2\sigma^2\right)\,,
\end{displaymath}
 where to get the canonical
normalization for the kinetic lagrangian one has to put
\begin{equation}
f_{\sigma}^{2}=\frac{8\,\mu_1\mu_2}{\pi^{2}}\frac{|\vec v_1||\vec
v_2| }{\,(|\vec v_1|+|\vec v_2|)} \label{couplingGoldston}\,.
\end{equation}

Using the result in (\ref{m}) one finds at the leading order: \be
m_\sigma^2=4\Delta^2\frac{m_1m_2}{\mu^2}\log\frac{\mu}{\Delta}\,,\ee
which coincides with the result of \cite{beane}. It should be
noted that in \cite{Schafer2} the masses of the NGB bosons in the
CFL case do not present any logarithmic enhancement factor. The
major difference between our approach and that of
\cite{Schafer2}, a part from the difference in the models that
are studied (CFL versus 2SC), is in the coupling between quarks,
as  we use a BCS four-fermion interaction while in \cite{Schafer2}
one considers one gluon exchange interaction.

Nonzero quark masses produce a change in the coupling constant
$f_\sigma$ and a reduction of the velocity of the pseudo-NGB in
the medium, i.e.\be v^2=\frac{|\vec v_1||\vec
v_2|}{3}=\frac{\sqrt{1-x_1^2}\sqrt{1-x_2^2}}{3} \label{1/3}\,,\ee
whereas for massless quarks $ v^2=1/3$. We wish to stress that
Eq. (\ref{1/3}) constitutes a new result which goes beyond the
expansion in $m/\Delta$ commonly used in the computation of mass
effects in color superconductivity.

\begin{center}
{\bf{Acknowledgements}} \end{center} We wish to thank M.
Mannarelli, K. Rajagopal and T. Sch\"afer for useful discussions.
Two of us, R.C. and G. N., wish to thank the DESY theory group,
where this work was completed, for the very kind hospitality.


\begin{thebibliography}{99}

\bibitem{Alford:1997zt}
M.~G.~Alford, K.~Rajagopal and F.~Wilczek,
Phys.\ Lett.\ B {\bf 422} (1998) 247 [arXiv:hep-ph/9711395].
%
\bibitem{Rajagopal:2000wf}
K. Rajagopal and F.Wilczek,in Handbook of QCD, M. Shifman ed.
(World Scientific 2001),
[arXiv:hep-ph/0011333].
\bibitem{Son:1999cm}
D.~T.~Son and M.~A.~Stephanov,
Phys.\ Rev.\ D {\bf 61} (2000) 074012
[arXiv:hep-ph/9910491].
%
\bibitem{Rho:1999xf}
M.~Rho, A.~Wirzba and I.~Zahed,
Phys.\ Lett.\ B {\bf 473} (2000) 126 [arXiv:hep-ph/9910550].
%
\bibitem{Hong:1999ei}
D.~K.~Hong, T.~Lee and D.~P.~Min,
Phys.\ Lett.\ B {\bf 477} (2000) 137 [arXiv:hep-ph/9912531].
%
\bibitem{beane}
S.~R.~Beane, P.~F.~Bedaque and M.~J.~Savage,
Phys.\ Lett.\ B {\bf 483} (2000) 131 [arXiv:hep-ph/0002209].


\bibitem{raku}
J.Kundu and K.Rajagopal, [arXiv:hep-ph/0112206].

\bibitem{Schafer2}
T.~Schafer,
Phys.\ Rev.\ D {\bf 65}, 074006 (2002) [arXiv:hep-ph/0109052].
%

\bibitem{Bedaque:2001je}
P.~F.~Bedaque and T.~Schafer,
Nucl.\ Phys.\ A {\bf 697} (2002) 802 [arXiv:hep-ph/0105150].


\bibitem{Kaplan:2001qk}
D.~B.~Kaplan and S.~Reddy,
Phys.\ Rev.\ D {\bf 65} (2002) 054042 [arXiv:hep-ph/0107265].

\bibitem{HDET}
D. K. Hong, Phys. Lett. {\bf B473} (2000) 118, {\tt
hep-ph/9812510}; D.~K. Hong, Nucl. Phys. {\bf B582} (2000) 451,
[arXiv:hep-ph/9905523];
 S.R. Beane, P.F. Bedaque and M.J. Savage, Phys. Lett.
{\bf B483} (2000) 131, [arXiv:hep-ph/0002209].

\bibitem{casa2}
R. Casalbuoni, R. Gatto and G. Nardulli, Phys. Lett. B {\bf 498}
(2001) 179 and Erratum-ibid. B517 (2001) 483,
[arXiv:hep-ph/0010321].

%
\bibitem{cini}
M. Cini and B. Touschek, Nuovo Cimento {\bf 7} (1958) 422.

\bibitem{Alford:2000ze} M.~G.~Alford,
J.~A.~Bowers and K.~Rajagopal,
Phys.\ Rev.\ D {\bf 63} (2001) 074016 [arXiv:hep-ph/0008208].

\bibitem{Bowers:2002xr}
J.~A.~Bowers and K.~Rajagopal,
Phys.\ Rev.\ D {\bf 66} (2002) 065002 [arXiv:hep-ph/0204079].


\bibitem{nard1}G. Nardulli, Riv. Nuovo Cimento, {\bf 25}, N. 3 (2003), [arXiv:hep-ph/0202037].



\end{thebibliography}
\end{document}